\documentclass[11pt,aps,prd,preprint,preprintnumbers,showpacs,showkeys,groupedaddress,floatfix]{revtex4}
\usepackage{graphicx}
\usepackage{epsfig}
\usepackage{dcolumn}
\usepackage{bm}
\usepackage{subfigure}
\usepackage{lscape}
\usepackage{amsmath}
\usepackage{amssymb}
\usepackage{setspace}
\usepackage{xcolor}
\usepackage[
           colorlinks,
           linkbordercolor={0 0 0},
           pdfborder={0 0 0},
           linkcolor=blue,
           citecolor=blue,
           urlcolor=blue,
           breaklinks=true]{hyperref}

\allowdisplaybreaks

\newlength{\x}
\newlength{\y}
\newlength{\z}
\urldef\milcurl\url{http://www.physics.utah.edu/~detar/milc/}

\phantomsection

\begin{document}

\title{Dynamical Restoration of $Z_N$ Symmetry in $SU(N)+$Higgs Theories
}

\author{Minati Biswal}
\email{mbiswal@imsc.res.in}
\author{Sanatan Digal}
\email{digal@imsc.res.in}
\author{P. S. Saumia}
\email{saumia@imsc.res.in}
\affiliation{The Institute of Mathematical Sciences,
Chennai, 600113, India}

\begin{abstract}
We study the $Z_N$ symmetry in $SU(N)+$Higgs theories with the Higgs field in the 
fundamental representation. The distributions of the Polyakov loop show that the $Z_N$ symmetry
is explicitly broken in the Higgs phase. On the other hand inside the Higgs symmetric phase
the Polyakov loop distributions and other physical observables exhibit the $Z_N$ symmetry.
This effective realization of the $Z_N$ symmetry in the theory changes the nature of the
confinement-deconfinement transition. We argue that the $Z_N$ symmetry will lead to time 
independent topological defect solutions in the Higgs symmetric deconfined phase which will 
play important role at high temperatures.
\end{abstract}

\maketitle

\section{Introduction}
\label{sec:intro} 


It is well known that most phenomena in pure $SU(N)$ gauge theories do not
depend on the representations of the gauge fields \cite{Halliday:1981te,Greensite:1981hw,
Gavai:1984ah, Kogut:1982rt,Ardill:1982gm,Datta:1997nv,Myers:2009df,Creutz:2007fe,Datta:1999np}. 
It is considered that both the fundamental and adjoint representations are equally valid 
representations of the non-abelian gauge fields and differences specific to representations 
are in general considered unphysical. 
The preference to a particular representation arises when the gauge fields are coupled 
to the matter fields. In the presence of the matter fields the two representations of 
the gauge fields are not equivalent. In quantum field theories such as the quantum 
chromodynamics $(QCD)$ and the electroweak $(EW)$ theory, which describe the strong and 
electro-weak forces of nature respectively, the matter fields are in the fundamental 
representations. The gauge invariance of these theories requires that 
the gauge fields also be in the fundamental 
representation. Given that there is a clear preference to the 
fundamental representation of the gauge fields, the physics aspects specific to this 
representation can play important role in these theories.

One of the important physics issue which arises in the fundamental representation is 
the $Z_N$ symmetry. At finite temperatures the gauge fields are periodic along the 
temporal direction \cite{Svetitsky:1985ye}. This boundary condition requires 
that in the temporal direction the gauge transformations are periodic up to a factor $z$, 
which is an element of the center (${Z}_N$) of the gauge group $SU(N)$. A
gauge transformation which is periodic upto a phase factor $z$ (in the temporal
direction)  non-trivially transforms the Polyakov loop $(L)$, which is the trace of a 
path ordered product of exponentials of the temporal gauge field $A_0$ along the shortest
temporal loop. The Polyakov loop picks up the element $z$ as 
a phase factor, i.e $L\rightarrow zL$ \cite{Svetitsky:1985ye}.
All possible gauge transformations of the Polyakov loop then form the $Z_N$ symmetry group. 
This symmetry plays an important role in the finite temperature confinement-deconfinement transition 
in pure $SU(N)$ gauge theories. In the deconfined phase the Polyakov loop acquires a non-zero 
expectation value which leads to the spontaneous breaking of the $Z_N$ symmetry. On the other 
hand in the confined phase it has zero expectation value. This property of the Polyakov loop 
across the confinement-deconfinement transition makes it an ideal candidate for an order 
parameter for this transition\cite{McLerran:1981pb}.

Even though the above non-periodic gauge transformations preserve the boundary conditions of the gauge 
fields they do not preserve the temporal boundary condition of the matter fields 
in the fundamental representation. After a gauge transformation for which $z \ne I$($I$ is 
the identity element of $Z_N$) bosonic(fermionic) matter fields are no more periodic(anti-periodic). 
These gauge transformations therefore can not act on the matter fields. 
However it still makes sense
to consider these $Z_N$ gauge transformations by restricting their actions only to the gauge 
fields. These transformations, which are not like the conventional gauge 
transformations acting both on the gauge and the matter fields, will not leave the action of
the full theory invariant. However a given gauge field configuration as well 
as it's $Z_N$ transformations are both valid configurations and  
will contribute to the partition function of the full theory. 
Their individual contribution to the partition 
function will decide the relative ``Boltzmann" probability of these two
configurations in a thermal ensemble. Even though the classical action does not have
the $Z_N$ symmetry ultimately the fluctuations of the fields will decide if the $Z_N$ symmetry is
relevant in presence of matter fields. Here by $Z_N$ symmetry we imply that the gauge 
transformations are acting only
on the gauge fields. The Higgs fields can be gauge transformed only when the gauge
transformations correspond to the identity of $Z_N$.   

The issue of $Z_N$ symmetry in the presence of fundamental matter fields has been extensively 
studied in the literature \cite{Gross:1980br,Weiss:1980rj,Weiss:1981ev,Ogilvie:2008fz,Myers:2008zm}. 
It was shown that the $1-loop$ perturbative effective potential for the Polyakov loop has meta-stable 
states with negative entropy \cite{Belyaev:1991np} in the presence of fermions. In these studies, 
however, only the zero mode of the Polyakov loop is coupled to the matter fields. Higher modes 
of the 
Polyakov loop, which actually give rise to the spontaneous breaking of the $Z_N$ symmetry, 
may resolve the problem of negative entropy. Subsequent studies using effective models 
\cite{Green:1983sd, Digal:2000ar} and lattice QCD studies \cite{Karsch:2001nf, Deka:2010bc} 
have shown that the presence of fermions acts as an external effective field on the Polyakov loop 
thereby breaking the $Z_N$ symmetry explicitly. 
Although there have been a lot of non-perturbative studies on 
the confinement-deconfinement transition of $SU(N)$ gauge theories coupled to fundamental bosonic 
fields \cite{Damgaard:1986jg,Damgaard:1987kk,Munehisa:1986jc} but very few have addressed the issue of
the $Z_N$ symmetry in these theories. 
In this work we carry out 
non-perturbative study of the $Z_N$ symmetry in the presence of bosonic matter fields in
the fundamental representation. 
More efforts are needed to address the issues related to the $Z_N$ symmetry in the presence of matter 
fields such as the thermodynamic properties of meta-stable states, strength of the symmetry breaking field 
etc. through higher order corrections to the effective 
potential and by non-perturbative Monte Carlo simulations. 

To study the $Z_N$ symmetry we focus mainly on the properties of the Polyakov loop as it 
is most sensitive to this symmetry. 
We compute the distribution of the Polyakov loop using the Monte 
Carlo simulations of the partition function. We have carried out simulations for the cases of 
$N=2$ and $N=3$. The distribution of the Polyakov loop is found to be similar to the distribution 
of the magnetization in the $N-$state Potts model (which has $Z_N$ symmetry) in the presence of 
the external field. The external field causes asymmetry in the distributions of the magnetization
which otherwise has the $Z_N$ symmetry. The larger the external field is larger is the asymmetry
in the distribution of the magnetization. In the present case the asymmetry of the Polyakov loop 
distribution is found to vary with the Higgs condensate. It is observed that the distribution
has large(small) asymmetry when the condensate is large(small). These results suggest that the 
external field for the Polyakov loop (the $Z_N$ symmetry) depends on the Higgs field. 
It is never expected that the external field vanish as long as there is interaction between
the gauge and the Higgs fields.
Surprisingly it is found that for a suitable choice of external parameters, when the system is 
in the Higgs symmetric phase, the Polyakov loop distribution exhibits the $Z_N$ symmetry. 
The simulation results also show that the different $Z_N$ states in the deconfined phase
have the same free energy. This implies the vanishing of the effective external field. This occurs 
while there is non-zero interaction (correlation) between the gauge and the Higgs fields. 
In this case the nature of the confinement-deconfinement transition is almost same as in the pure 
gauge case. Apart from affecting the confinement-deconfinement
transition the $Z_N$ restoration in the theory will lead to 
presence of domain walls and strings defects ($N>2$) at very high temperatures in the deconfined 
phase. Previously the effective potential calculations have 
shown that the $Z_N$ symmetry is restored only in the limit of infinitely heavy 
Higgs mass, that is basically when the Higgs field decouples from the gauge fields. In contrast in our 
non-perturbative studies the $Z_N$ symmetry is realized even when the Higgs has finite mass and 
its interaction with the gauge fields is non-zero. It would be interesting to investigate this 
symmetry in the presence of fundamental fermion fields in view of its restoration in the
presence of the Higgs field. We mention here that conventionally symmetry restoration means
that the distribution of the order parameter (the Polyakov loop in the present context) is symmetrically
peaked around zero. In the present context by symmetry restoration we imply that the full theory 
exhibits the corresponding symmetry.

The paper is organized as follows. In the following in section-II we discuss the $Z_N$ 
symmetry in $SU(N)+$Higgs theories. In section-III we present our numerical simulations and results. 
In section-IV we present our discussions and conclusions.

\section{The $Z_N$ symmetry in the presence of fundamental Higgs fields} 

The Euclidean $SU(N)$ action for the gauge fields $A^a_\mu$ ($a=1,2,...N^2-1$)
in the fundamental representation is given by,
\begin{equation}
S = \int_V d^3x\int_0^\beta d\tau 
\left\{ {1 \over 2} Tr\left(F^{\mu\nu}F_{\mu\nu}\right) \right \}.
\end{equation}
$V$ is spatial volume and $\beta$ is the extent in temporal direction. The
gauge field strength $F_{\mu\nu}$ is given by
\begin{equation}
F_{\mu\nu}=\partial_\mu A_\nu-\partial_\nu A_\mu + g[A_\mu,A_\nu], ~~~~~~A_\mu=A^a_\mu T^a.
\end{equation}
The $N\times N$ matrices $T^a$'s are the generators of the $SU(N)$ gauge group. g is the gauge coupling 
constant. In the
Euclidean theory the gauge fields $A^a_\mu$ are periodic in the temporal direction, i.e 
$A^a_\mu(\vec{x},0)=A^a_\mu(\vec{x},\beta)$. Under a gauge transformation $U(\vec{\bf x},\tau) \in SU(N)$
the gauge fields transform as
\begin{equation}
A_\mu \longrightarrow U A_\mu U^{-1} + {1 \over g}\left(\partial_\mu U\right) U^{-1}. 
\end{equation}
Though the gauge fields must be periodic the gauge transformations $U(\vec{\bf x},\tau)$ need not be 
periodic in the temporal direction. The invariance of the pure gauge action and the periodicity of the
gauge fields both can be satisfied by gauge transformations which are periodic up to a factor
$z$ such as,
\begin{equation}
U(\vec{x},\tau=0)=z U(\vec{x},\tau=\beta).
\end{equation}
Where $z\in Z_N$ and $Z_N$ is the center of the gauge group 
$SU(N)$ \cite{McLerran:1981pb, Svetitsky:1982gs}. 
The Polyakov loop $(L)$ which is the path ordered product of links in the temporal direction,
\begin{equation}
L(\vec{\bf x})={1 \over N} Tr\left\{P e^{\left(-ig\int_0^\beta A_0 d\tau\right)}\right\}
\end{equation}
transforms as  $L \longrightarrow z L$ under a gauge transformation (Eq.(3)) with the boundary 
condition Eq.(4). 
Consequently the Polyakov loop behaves like a $Z_N$ spin and plays the role of an order 
parameter for the pure gauge confinement-deconfinement transition. Note that $L$ is the trace of an 
$SU(N)$ matrix. For $N=2$ the range of values $L$ can take is $\left [-1,1\right ]$. For
$N > 2$ it can take any value in a $n-$polygon in the complex plane whose vertices
are given by $e^{i{2\pi n \over N}},n=0,1,N-1$.

The modified action which describes the interaction of the gauge fields and the 
Higgs field $\Phi$ is given by,
\begin{equation}
S = \int_V d^3x\int_0^\beta d\tau 
\left\{ {1 \over 2} Tr\left(F^{\mu\nu}F_{\mu\nu}\right) + {1 \over 2} |D_\mu\Phi|^2
+{m^2 \over 2} \Phi^\dag\Phi + {\bar{\lambda} \over 4!}(\Phi^\dag\Phi)^2\right \}.
\end{equation}
The $\Phi$ field is a $N \times 1$ column matrix with complex elements. 
$m$,$\bar{\lambda}$ are the bare mass and the self-interaction strength of the $\Phi$ field
respectively.  The covariant derivative $D_\mu\Phi$ is defined as 
$D_\mu\Phi=\partial_\mu\Phi+ ig A_\mu\Phi$. 
Being a bosonic field, $\Phi$ satisfies periodic boundary condition in the temporal direction, i.e  
$\Phi(\vec{x},0)=\Phi(\vec{x},\beta)$. Under a gauge transformation $U(\vec{\bf x},\tau)$ the 
$\Phi$ field transforms as,
\begin{equation}
\Phi^\prime = U \Phi.
\end{equation}
It is obvious that $\Phi^\prime$ is periodic only when
the gauge transformations are periodic. Therefore the gauge transformations which are not periodic
are not allowed to act on the matter fields. Thus $Z_N$ group is not a symmetry of the classical 
action (Eq.6). However the actual manifestation of the $Z_N$ symmetry can be seen only after the
fluctuations of the gauge and matter fields are included as fluctuations play dominant role in these 
theories. The change in the action due $Z_N$ transformation acting only on the gauge fields can be 
compensated by fluctuations of the Higgs field. This leads to the complete realization/restoration 
of the $Z_N$ symmetry. In the following we describe the numerical Monte Carlo 
simulations and results.

\section{Simulations of the $SU(N)+$Higgs model}

In the Monte Carlo simulations of $SU(N)+Higgs$ model, 
the $4-$dimensional Euclidean space is replaced by a discrete lattice. The 
lattice sites are represented by $n=(n_1,n_2,n_3,n_4)$ where $n_i$'s are integers. 
The gauge field $A_\mu$ is replaced by the link variables 
$U_\mu = exp(-iagA_\mu$), where $a$ is the lattice constant/spacing. The link variable $U_\mu(n)$ lives on
the link between the sites $n$ and $n+\hat{\mu}a$, where $\hat{\mu}$ is a unit vector in the $\mu$th
direction. The Higgs field $\Phi(n)$ lives on the lattice site $n$. 
The discretized lattice action is given by,

\begin{equation}
S = \beta \sum_p {1 \over 2}Tr(2-U_p-U^\dag_p) - \kappa \sum_\mu Re\left [(\Phi^\dag_{n+\mu}U_{n,\mu}\Phi_n)\right]
+{1 \over 2} \left(\Phi^\dag_n\Phi_n\right) 
+ \lambda\left({1 \over 2} \left(\Phi^\dag_n\Phi_n\right) -1\right)^2
\end{equation}

where $U_p$ is the product of links in an elementary square $p$ on the lattice. The $\Phi$ field and other parameters 
are all dimensionless in the discretized action \cite{Kajantie:1995kf}. 
The Polyakov loop $L(n_i)$ at a spatial site $n_i$ is trace of the path ordered product of all temporal link
variables on the temporal loop going through $n_i$. A $Z_N$ rotation can be carried out by multiplying 
all temporal links on a fixed temporal slice of the lattice by an element of the $Z_N$ group. 
This operation leaves all terms of the above action invariant except the $\kappa$ dependent
term. This term is solely responsible for the explicit breaking of the $Z_N$ symmetry.

In the simulations an initial configuration of $\Phi_n$ and $U_{\mu,n}$ is selected. This initial configuration
is then repeatedly updated to generate a Monte Carlo history. In an update a new configuration is generated from an 
old one according to the Boltzmann probability factor $e^{-S}$ and the principle of detailed balance. 
These conditions are implemented using pseudo heat-bath algorithm for the $\Phi$ field \cite{Bunk:1994xs} and
the standard heat-bath algorithm for the link variables $U_\mu$'s
\cite{Creutz:1980zw,Cabibbo:1982zn}. Apart from updating procedure over relaxation methods are also used to reduce the 
autocorrelations between adjacent configurations along the Monte Carlo trajectory \cite {Whitmer:1984he}. 

\begin{figure}[h]
  \centering
  \subfigure[]
  {\rotatebox{360}{\includegraphics[width=0.45\hsize]
      {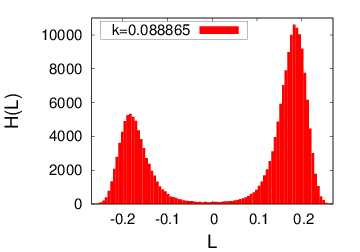}}
    \label{histo_Im_L_1a}
  }
  \subfigure[]
  {\rotatebox{360}{\includegraphics[width=0.45\hsize]
      {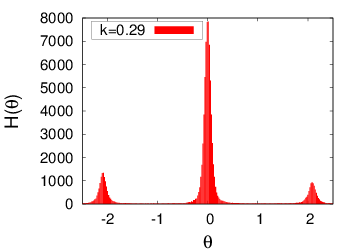}}
    \label{histo_Im_L_1b}
  }
  \caption{Distribution of Polyakov loop in the Higgs broken phase for (a) $SU(2)$, $16^3\times 4$ lattice and
(b) $SU(3)$, $8^3\times 4$ lattice.}
\end{figure}

The simulations are carried out for different values of $\beta,\kappa$ and $\lambda$. The coupling $\lambda$ controls 
the nature of the Higgs transition. The transition is first oder(crossover) for small(large) values of $\lambda$.
For a fixed $(\lambda,\beta)$ the parameter $\kappa$ plays the role of the transition parameter for the Higgs transition. 
For high $\kappa(\kappa>\kappa_c)$ the system is found to be in the Higgs phase with a non-zero Higgs condensate. 
With decrease in $\kappa$ the condensate starts to melt and at the critical point $\kappa=\kappa_c$ 
the system undergoes transition to the Higgs symmetric phase. For $\kappa < \kappa_c$ the Higgs condensate vanishes. 
For our purpose it suffices to fix the coupling $\lambda$ and study the $Z_N$ symmetry at various values of
$\kappa$. Given a ($\lambda$,$\kappa)$ small(large) $\beta$ corresponds to the 
confinement(deconfinement) phase. The confinement-deconfinement transition takes place at the critical point
$\beta=\beta_c$ \cite{Damgaard:1986jg,Damgaard:1986qe,Evertz:1986af,Damgaard:1987kk,Munehisa:1986jc}. 
To study the $Z_N$ symmetry at different $\kappa$
we compute the Polyakov loop distribution and simulate confinement-deconfinement transition. We also compute various
observables which are sensitive to the $Z_N$ symmetry. In Fig.1a we show the Polyakov loop 
distribution($H(L)$) in the deconfined phase for $N=2$ for $\lambda=0.005$ and $\kappa=0.088865$.
The explicit breaking of $Z_2$ symmetry is clearly seen in the distribution $H(L)$. The local maximum here corresponds to 
the meta-stable state of the system. For $N\ge 3$ the Polyakov loop is 
complex. For better illustration we show the distribution of phase of the Polyakov 
loop $H(\theta)$ instead of $H(L)$ on the complex plain. In Fig.1b we show $H(\theta)$ for $\lambda=0.1$ and 
$\kappa=0.29$ for $N=3$. The peak at $\theta=0$ clearly dominates the other two local maxima as a 
result of the $Z_3$ explicit symmetry breaking. It has been observed that the asymmetry in the above 
distributions increases when $\kappa$ is increased further. Beyond some value of $\kappa$ (which depends
on $\lambda$ and $N$) the local maxima(the meta-stable states) disappear.   

\begin{figure}[h]
  \centering
  \subfigure[]
  {\rotatebox{360}{\includegraphics[width=0.45\hsize]
      {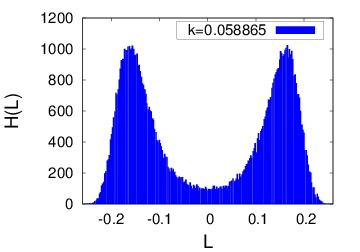}}
    \label{histo_Im_L_1a}
  }
  \subfigure[]
  {\rotatebox{360}{\includegraphics[width=0.45\hsize]
      {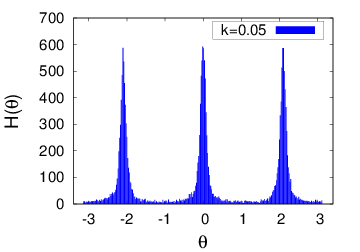}}
    \label{histo_Im_L_1b}
  }
  \caption{Distribution of Polyakov loop in the Higgs symmetric phase for (a) $SU(2)$, $16^3\times 4$ lattice and
(b) $SU(3)$, $8^3\times 4$ lattice.}
\end{figure}

The $Z_N$ symmetry is supposed to be there only when $\kappa=0$ as the matter and gauge fields 
decouple. Surprisingly it is found in our simulations that in the Higgs symmetric phase ($0< \kappa<\kappa_c$) 
the distributions of the Polyakov loop exhibit the $Z_N$ symmetry. 
This is evident in the distribution $(H(L))$ of the Polyakov loop for $N=2$ shown in Fig.2a. Similarly 
the distribution $H(\theta)$ for $N=3$ shows the $Z_3$ symmetry. For small $\kappa$ the Higgs correlation 
length can become shorter than the lattice spacing, i.e $\Phi_n$ and $\Phi_{n+\mu}$ are not correlated. With
the product $\Phi_n\Phi^{\dag}_{n+\mu}$ having no preferential orientation with respect to $U_\mu(n)$ 
the $\kappa$ term in Eq.(8) can not affect the $Z_N$ symmetry. Though this is plausible but our
simulations suggest that this is not the reason for the $Z_N$ realization/restoration. The $\kappa$ term
was found to be non-zero finite. The product $\Phi_n\Phi^\dag_{n+\mu}$ tend to align with $U_\mu(n)$. 
When a $Z_N$ rotation ($(\Phi,U) \rightarrow (\Phi,U_g)$) is carried out on any configuration from the thermal
ensemble the resulting configuration is found to be out of equilibrium. This is because the new
configuration has far higher action (Eq.(8)) then any configuration in the thermal ensemble. Interestingly
this cost in the action can be compensated by varying the $\Phi$ field, i.e $\Phi \rightarrow \Phi^\prime$,
coupled with the gauge rotation of the links. $\Phi^\prime$ can be obtained by Monte Carlo updates of
$\Phi$, though it is not clear how $\Phi$ and $\Phi^\prime$ are related. We observed that the symmetry 
$(\Phi,U)\rightarrow (\Phi^\prime,U_g)$ is there only in the Higgs symmetric 
phase ($\kappa < \kappa_c$) and when the number of lattice points in the temporal direction is $N_\tau \ge 4$.

To see the $Z_N$ symmetry in the Polyakov loop distribution, the tunneling between the different $Z_N$
sectors has to be high. The tunneling rate decreases away from the transition point and also for larger
lattice size. For these cases even for a reasonably large statistics it is unlikely that the population of the 
different Polyakov loop sectors will be found same. For example, for $\beta=2.38$ and $16^3\times 4$ lattice
we do not see any tunneling between the different $Z_2$ sectors up to $2\times 10^6$ statistics.
However the histogram of the Polyakov loop in the two sectors are in perfect agreement when one
distribution is $Z_2$ rotated as is seen clearly in Fig.3a. Apart from the Polyakov loop distributions
we also compute the free energy of the different Polyakov loop sectors. In Fig.3b we show the average 
value of the gauge action vs $\beta$ for the two $Z_2$ states(called $+$ve and $-$ve) for $N=2$. The
gauge action for the $+$ve($-$ve) sector is calculated by taking the average over configurations for
which the Polyakov loop is $+$ve($-$ve). The gauge actions for the two $Z_2$ states are identical for all 
$\beta$. The free energy of each of these states can now be computed by integrating the gauge action 
$S_G(\beta)$ in $\beta$\cite{Boyd:1995zg,Boyd:1996bx}. Since the gauge action are identical, the free
energy will be same for the two Polyakov loop sectors.

\begin{figure}[h]
  \centering
  \subfigure[]
  {\rotatebox{360}{\includegraphics[width=0.48\hsize]
      {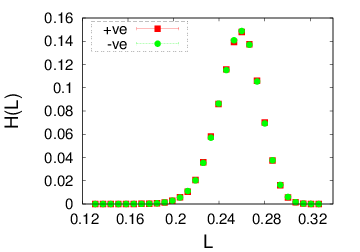}}
    \label{histo_Im_L_3a}
  }
  \subfigure[]
  {\rotatebox{360}{\includegraphics[width=0.48\hsize]
      {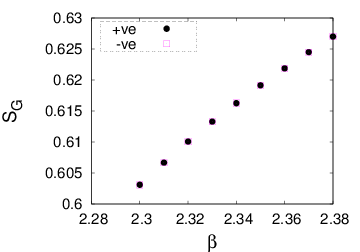}}
    \label{histo_Im_L_3b}
  }
  \caption{ (a) Comparison of Polyakov loop distributions ($\beta=2.38,\lambda=0.005,\kappa~0.056$, 
lattice = $16^3\times 4$) and (b) Gauge action $S_G(\beta)$ for the two Polyakov loop sectors for $N=2$. } 
\end{figure}

The confinement-deconfinement transition for $N=2$ for small $\kappa$ has been investigated previously 
\cite{Damgaard:1986jg,Damgaard:1986qe,Evertz:1986af,Damgaard:1987kk,Munehisa:1986jc}. These studies have shown 
that the average value of the Polyakov loop does have critical behavior and 
found to be in the universality class of the Ising model. In this study for
the first time we carry out the finite size scaling analysis of the Binder cumulant \cite{Binder:1981sa}.
In Fig.4a the Binder Cumulant \cite{Binder:1981sa} around transition point is shown for different spatial 
volumes. The value of the Binder Cumulant at the crossing point corresponds to the universality class of the 
$3-$D Ising model. Further the scaling of the Binder Cumulant, shown in Fig.4b, gives a value for the 
critical exponent $\nu~0.62998$ which is also consistent with the same universality class. These
results clearly show that the confinement-deconfinement transition is second order even for finite but
small $\kappa$. Conventionally it is thought that the 
confinement-deconfinement transition is true second order only for $\kappa=0$. 
We believe that the origin of this second order confinement-deconfinement transition at $\kappa \ne 0$ is 
because the fluctuations respect the $Z_2$ symmetry. The realization of the $Z_2$ symmetry and the critical 
behavior of the Polyakov loop for finite $\kappa$ suggests that there should be a line of second order 
confinement-deconfinement transitions starting from $\kappa=0$ line on the phase diagram. 

\begin{figure}[h]
  \centering
  \subfigure[]
  {\rotatebox{360}{\includegraphics[width=0.48\hsize]
      {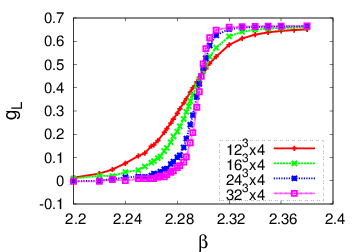}}
    \label{histo_Im_L_4a}
  }
  \subfigure[]
  {\rotatebox{360}{\includegraphics[width=0.48\hsize]
      {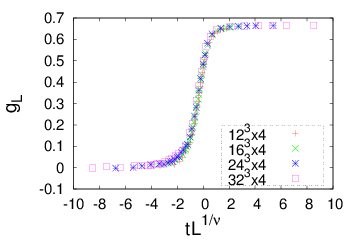}}
    \label{histo_Im_L_4b}
  }
  \caption{ (a) Binder Cumulant and (b) its scaling for $SU(2)$. } 
\end{figure}

\section{Discussions and Conclusions}

We have studied the $Z_N$ symmetry in $SU(N)+$Higgs theories for $N=2,3$ using numerical Monte
Carlo simulations. The presence of the Higgs fields explicitly breaks the $Z_N$ symmetry 
which is reflected in the asymmetry in the Polyakov loop distribution.
The strength of the explicit symmetry breaking 
varies with the parameters $\lambda$ and $\kappa$. On the other hand, given a $(\lambda,\kappa)$ 
the strength does not vary much with the confinement-deconfinement transition parameter $\beta$. 
The patterns of explicit symmetry breaking observed in $N=2$ and $N=3$ are very similar. This suggests 
that this pattern will continue to hold for higher $N$.

The explicit breaking of $Z_N$ symmetry has clear pattern along any trajectory on $\lambda-\kappa$ 
plane of decreasing $\kappa$ and the Higgs condensate. It has been observed that for large values of 
these variables the explicit symmetry breaking is so large that $H(L)$ and $H(\theta)$ have only one 
peak in deconfined phases. The $Z_N$ symmetry is maximally broken in this case. Further down as $\kappa$
and the Higgs condensate decrease multiple peaks in the distributions do appear in 
the deconfined phase. For some other trajectories on the $\lambda-\kappa$ plane , in the Higgs phase region, 
it is possible that only one of these two situations may arise. As the trajectory crosses the Higgs 
transition point $\kappa_c$ the explicit symmetry breaking drops sharply. Close to the transition point 
in the Higgs symmetric phase $H(L)$ and $H(\theta)$ peaks are almost degenerate. It will be important
to see the effect of $N_\tau$ (number of lattice points in temporal direction) on this small but finite
explicit symmetry breaking. It is possible that the explicit symmetry breaking vanishes in all of the Higgs 
symmetric phase in the infinite volume limit. 

Conventionally it is expected that the explicit symmetry breaking will vanish only when
$\kappa$ is zero. In our simulations (with $N_\tau=4$) it is found that the explicit symmetry breaking 
vanishes in the Higgs symmetric phase away from the transition point. The value of $\kappa$ for
which the symmetry restored in the theory occurs depends on $\lambda$. For larger $\lambda$ the
the restoration of the $Z_N$ symmetry occurs at a higher value of $\kappa$. This suggests
that for a given $\beta$ a line divides the $\lambda-\kappa$ plane into $Z_N$ symmetric
and $Z_N$ broken regions. In the $Z_N$ symmetric region the $Z_N$ symmetry is spontaneously 
broken for $\beta > \beta_c$ which leads to $N$ degenerate states. 
All physical observables such as the gauge action, the 
kinetic term etc. are found to be same for all the $Z_N$ states. As a consequence the free energies 
of the different $Z_N$ states are the same. Our results clearly indicate that the Higgs condensate 
plays role of the $Z_N$ symmetry breaking field. However more work is needed to relate the 
Higgs condensate to the effective field for the $Z_N$ symmetry. In this work we have used the Higgs 
transition point to infer the values of the Higgs condensate. Since the Higgs field is not gauge 
invariant the Higgs condensate is not well defined. We plan to calculate the Higgs condensate by
appropriately choosing a gauge which will make the Higgs condensate well defined and find out the
connection between the Higgs condensate and the explicit symmetry field for $Z_N$.

The realization of $Z_N$ symmetry at non-zero $\kappa$ is in contradiction with effective potential 
calculations which show that the $Z_N$ symmetry will be restored only when the Higgs mass is 
infinite. In these calculations only the zero mode of the Polyakov loop is 
coupled to the matter fields. We expect that taking care of the higher modes of the Polyakov 
loop will reduce the discrepancy between the non-perturbative and analytic approaches.
The restoration of the $Z_N$ symmetry in the Higgs symmetric phase has important implications 
for the phase diagrams of $SU(N)+$Higgs theories. For $N=2$ previously the 
confinement-deconfinement transition was thought to be a crossover for non-zero $\kappa$. 
Our results show that there will be a line
of second order confinement-deconfinement transitions in the $\beta-\kappa$
plane extending from the point $(\beta_c(\kappa=0),\kappa=0)$. Since the $Z_N$ symmetry is 
spontaneously broken at high temperatures in the Higgs symmetric phase with
vanishing condensate it will lead to rich structures in this phase. Spontaneous symmetry breaking
of the $Z_N$ symmetry will lead to time independent topological defects solutions such as 
domain walls, strings etc. These defects can form even when the $Z_N$ 
symmetry is mildly broken but they are not time independent and are short 
lived. We mention here that the restoration of $Z_N$ symmetry may be possible in the case of
gauge fields coupled to fundamental fermions as well. In this case 
$\bar{\psi}\gamma_0\psi$ (which couples to the $A_0$ field) may play the role similar to the
Higgs field in restoring the $Z_N$ symmetry.

\acknowledgements{We thank Saumen Datta for important comments and suggestions. We thank Mridupawan Deka 
for useful discussions and for providing us with the MPI code.}

\end{document}